\documentstyle[12pt,aaspp]{article}
\tighten
\def\refindent{\par\noindent\parskip=4pt\hangindent=3pc\hangafter=1 }
\def\apj#1#2#3{\refindent#1,  {ApJ, }{#2}, #3}

\def\aa#1#2#3{\refindent#1,  {A\&A, }{#2}, #3.}
\def\aas#1#2#3{\refindent#1,  {A\&AS, }{#2}, #3.}

\def\spose#1{\hbox to 0pt{#1\hss}}

\def\eg{{\it e.g.\ }}

\def\lta{\mathrel{\spose{\lower 3pt\hbox{$\mathchar"218$}}
     \raise 2.0pt\hbox{$\mathchar"13C$}}}
\def\gta{\mathrel{\spose{\lower 3pt\hbox{$\mathchar"218$}}
     \raise 2.0pt\hbox{$\mathchar"13E$}}}

\font\tenmib=cmmib10 \textfont"E=\tenmib
\font\tenbsy=cmbsy10 \textfont"F=\tenbsy

\def\Oma{\Omega_{\rm max}}
\begin{document}
\hfill{\sl To be published in the Astrophysical Journal}
\par

\title{FAST ROTATION OF NEUTRON STARS}

\author{Jean--Pierre LASOTA \altaffilmark{1},
Pawe{\l} HAENSEL \altaffilmark{2},
\and Marek Artur ABRAMOWICZ \altaffilmark{3}}

\altaffiltext{1} {UPR 176 du CNRS, DARC, Observatoire de Paris, Section de
Meudon,
 92195 Meudon Cedex, France (lasota@obspm.fr)}

\altaffiltext{2} {N. Copernicus Astronomical Center, Polish Academy of
Sciences,
Bartycka 18, 00-716 Warszawa, Poland (haensel@camk.edu.pl)}

\altaffiltext{3}{Department of Astronomy \& Astrophysics,
University of Gothenburg and Chalmers University of
Technology, 412 96 Gothenburg, Sweden (marek@fy.chalmers.se)}

\begin{abstract}
We show that  for  realistic equations of state of dense matter,
the universal proportionality factor
relating the maximum rotation rate of neutron stars due to
mass-shedding limit to the mass and radius of maximum allowable mass
configuration of non-rotating models results from a universal
proportionality between masses and radii of static maximum-mass neutron stars
and those of maximally rotating configurations.
These empirical relations cannot
be obtained in the slow rotation approximation.
\end{abstract}

\keywords{ pulsars--stars: neutron--stars: rotation}

\section{INTRODUCTION}

The question of how fast a pulsar can spin has  recently been discussed by
several authors (see \eg Friedman \& Ipser 1992, Cook et al. 1994a, b).
Independently of the
possible mechanisms of pulsar spin-up, for any particular equation
of state there is always an upper limit for the final allowed
rotational speed, $\Omega_{\rm max}$. The limit is determined either
by the mass shedding condition, {\it i.e.} that the angular velocity
of the configuration, stable with respect to axisymmetric perturbations,
 equals the Keplerian velocity at the surface, or by the
condition of the onset of non-axisymmetric (e.g. gravitational radiation
reaction)  instabilities. In this paper
we discuss the limit given by the mass shedding condition.

Haensel and Zdunik (1989; hereafter HZ) noticed that for realistic
 equations of state of dense matter,
  the numerically calculated values of the shedding
limit $\Omega_{\rm max}$ can be fitted, with an accuracy better than
5\%, by an empirical formula
$$
\Oma = {\cal C}_{\Omega}\left ({GM_{\rm max}}\over {R_{\rm
                                     max}^3}\right )^{1/2}, \eqno(1.1)
$$
where $M_{\rm max}$ is the maximal mass of the non-rotating neutron
stars with the same equation of state, $R_{\rm max}$ is the radius
corresponding to $M_{\rm max}$, and ${\cal C}_{\Omega}$ is a dimensionless
phenomenological constant, independent of the equation of state.
$\Omega_{\rm max}$ is an angular velocity of rigid rotation as
measured by a
stationary observer at infinity. HZ determined that the best fit is
for ${\cal C}_{\Omega} = 0.67$.

The calculations of Friedman, Ipser  and Parker (1989),
performed  for a very broad set of equations of state, yielded
${\cal C}_\Omega=0.66$, quoted in Friedman (1989), Friedman and
Ipser (1992). It should be mentioned, that the value of ${\cal
C}_\Omega=0.62$, quoted in the original paper of Friedman, Ipser
and Parker (1989), was actually a very rough estimate of ${\cal
C}_\Omega$ (J.L. Friedman, private communication).
 Calculations based on recent very accurate numerical methods are
  in good
agreement with the original HZ choice of $C_\Omega$.  For example, values of
$\Omega_{\rm max}$ calculated for several equations of state  by
Lattimer et al. (1990) differ from  the HZ version
of the empirical formula by less than 4\%.
  Most recent calculations
in the full framework of General Relativity by  Salgado et al. (1994a,b;
hereafter SBGH)
based on the spectral methods (Bonazzola et al. 1993) show that
configurations with causal equations of state (EOS) satisfy (1.1) with
${\cal C}_{\Omega} =0.67$ with an accuracy of better than 5\%
(Haensel, Salgado and Bonazzola 1995, hereafter HSB).
 Results
obtained for causal EOS by Cook et al. (1994b) lead to very
similar  ``best fit" value of ${\cal C}_\Omega$.
  HSB found that the empirical  relation with ${\cal C}_\Omega=0.67$
fails for configurations constructed with non-causal equations of
state, where the sound speed, $(\partial P/\partial \rho)^{1/2}_{
S}$, may be greater than the speed of light within the neutron
star models.

 Empirical relations of universal character, valid for a broad
 range of realistic equations of state of dense matter, can
obtained also  for
other bulk parameters of neutron star models. For
example, a simple and surprisingly good
universal relation, connecting the maximum moment
of inertia for slow, rigid rotation, $I_{\rm max}$, to the mass and
radius of a static configuration with maximum allowable mass
(which is different from that with $I_{\rm max}$ !) was pointed
out by Haensel (1992). The existence of such universal empirical
relations is of practical importance: for any realistic EOS,
 empirical formulae enable rapid and still quite precise
 calculation of the upper bounds on global parameters of neutron
stars from the easily calculated parameters of the static
configuration with  maximum allowable mass. A different type
of a universal formula was obtained by Ravenhall and Pethick
(1994). Their formula, valid for a broad range of realistic
equations of state of dense matter, and useful for all except
lightest neutron stars, expresses  the moment of inertia in
terms of stellar mass and radius.

%The existence of a simple correlation between the maximum rate of
%rotation and the properties of the maximum-mass non-rotating
%configuration seems to indicate the presence of some underlying, very
%general, properties of rotating bodies in General Relativity.

The few attempts to explain the empirical relation, Eq. (1.1),
 were not concluded with
satisfactory results. The most elaborate discussion was presented by
Weber and Glendenning (1991; 1992).  They used numerical models of
rotating neutron stars calculated in the slow rotation approximation
of Hartle and Thorne (1968) to show that these also obey the
empirical
formula, albeit with ${\cal C}_{\Omega} \approx 0.75$.
 Although this is
obviously an important result, still lacking is a clear physical
understanding of the problem.

In a recent work Glendenning and Weber (1994) derived a
formula which  relates $\Omega_{\rm max}$
to  $M_{\rm max}/ R_{\rm max}^3 $ , in the slow
rotation approximation, where  $M_{\rm max}$ and $R_{\rm max}$
correspond to rotating configurations, but they do not discuss the
connection between this relation and the formula (1.1).
As we shall see below the slow rotation
approximation cannot be applied consistently to maximally rotating
neutron stars, as pointed out already by Hartle and Thorne (1968).

\section{THE EMPIRICAL RELATIONS BETWEEN
NON-ROTATING AND MAXIMALLY ROTATING CONFIGURATIONS}

One can see that, to a good approximation,
 the ``constant" ${\cal C}_{\Omega}$ is in fact
a slowly varying function of one variable, characterizing the
 EOS of dense matter, ${\cal C}_\Omega{\rm (EOS)}$ (Table 1 and
 HSB).
  Indeed Fig. 1 shows the
values of ${\cal C}_{\Omega}$ as given by eq. (1.1) for 12
 maximally rotating
models calculated by SBGH, as a function of the parameter
%%%%%%%%%%%%%%%%%%%%%%%%%%%%%%%%%%%%%%%%%%%%%%%%%%%%%%%%%%%%%%%%%%%%%%%%%%%%%%
$$ x_{\rm s} = {2 G M_{\rm max}\over R_{\rm max} c^2}
                \eqno (2.1)  $$
%%%%%%%%%%%%%%%%%%%%%%%%%%%%%%%%%%%%%%%%%%%%%%%%%%%%%%%%%%%%%%%%%%%%%%%%%%%%%%
for static maximum mass configurations.
It is clear that ${\cal C}_\Omega({\rm EOS})$
is slowly varying, and for the range of
parameters of interest, can be well approximated by
a monotonic function of $x_{\rm s}$.
%This fact
%appears even clearer if one considers series of polytropic models
%(see Table 2, based on HSB and Salgado (1994)).
If one restricts to realistic EOS, which
are both causal and stiff enough to support observed masses of pulsars, the
range of relevant $x_{\rm s}$ becomes rather narrow and the approximation
of ${\cal C}_{\Omega}({\rm EOS})$ by a constant is
quite satisfactory (see HSB).

It is well known that for realistic EOS,
 rotation increases the value of maximum mass
of a neutron star by about 20$\%$. We find however that,
with a very good approximation,
there exists
a {\it universal} relation between the maximum mass of non-rotating neutron
star configuration and the mass of a configuration rotating with
an angular speed $\Oma$. The relation is:
%%%%%%%%%%%%%%%%%%%%%%%%%%%%%%%%%%%%%%%%%%%%%%%%%%%%%%%%%%%%%%%%%%%%%%%%%%%%%
$$M_{\rm max}({\rm rot}) = {\cal C}_M M_{\rm max}({\rm stat}) ,
\eqno (2.2) $$
%%%%%%%%%%%%%%%%%%%%%%%%%%%%%%%%%%%%%%%%%%%%%%%%%%%%%%%%%%%%%%%%%%%%%%%%%%%%%
where  $ M_{\rm max}({\rm rot})$ and $ M_{\rm max}({\rm stat})$ are
respectively the
mass of the configuration in maximum rotation and the maximum mass of the
static neutron star for the same equation of state. Of course,
 a similar relation holds trivially for any  specific EOS, and
yields a specific value of the proportionality constant,
${\cal C}_M({\rm EOS})$. The values of
${\cal C}_M ({\rm EOS}) $, calculated for a broad set of EOS using the results
 of  SBGH,  are given in Table 1. The best fit of relation (2.2) to
numerical results of SBGH is obtained for ${\cal C}_M=1.18$.
Then, relation (2.2) reproduces exact results within better
than 3\% (Fig.2).

A similar  relation is found for the radii of the corresponding
configurations:
%%%%%%%%%%%%%%%%%%%%%%%%%%%%%%%%%%%%%%%%%%%%%%%%%%%%%%%%%%%%%%%%%%%%%%%%%%%%%
$$R_{\rm max}({\rm rot}) = {\cal C}_R  R_{\rm max}({\rm stat}) ,
\eqno (2.3) $$
%%%%%%%%%%%%%%%%%%%%%%%%%%%%%%%%%%%%%%%%%%%%%%%%%%%%%%%%%%%%%%%%%%%%%%%%%%%%%
where  $R_{\rm max}({\rm rot})$ and $R_{\rm max}({\rm stat})$
are the radii
of the maximally rotating and the static configurations respectively.
 The best fit value of a {\it constant}
${\cal C}_R$ in relation (2.3), based on numerical results of
SBGH for causal EOS, is ${\cal C}_R=1.34$. Relation (2.3) reproduces
then exact results for {\it causal} EOS within better than 4\%
(Fig. 3).

As in the case of relation (2.2), one can
introduce the factor  ${\cal C}_R ({\rm EOS})$ (Table 1). The
dependence of ${\cal C}_R$ on the EOS can be well approximated
by a monotonically decreasing function of $x_{\rm s}$ - this is
visualized in Fig. 4. No such a trend was found for
${\cal C}_M{\rm (EOS)}$), Fig. 5.

Our analysis have been based on a set of numerical results,
obtained for twelve realistic EOS in SBGH. We
have restricted to the SBGH set in order to keep the homogeneity
of the sample of numerical results. Numerical results obtained
by other authors  (Friedman, Ipser \& Parker 1989, Cook et al.
1994b)  are found to be slightly different from those of SBGH.
%CHANGE: next sentence has been changed
 This is due to the fact, that the precise
determination of the maximally rotating
configuration -- which requires very high precision of numerical
procedure -- turns out to be sensitive to such details as, e.g.,
the interpolation method used for the determination of the maximum
frequency model
  (Stergioulas \& Friedman  1995, E. Gourgoulhon, private communication).
 However, uncertainties resulting from this dependence on the
specific sample of numerical results obtained for realistic
EOS, are consistent with the precision of relations (2.2), (2.3),
discussed above.

One can easily see that the empirical  ${\cal C}_\Omega$
constant may be obtained, within a very good approximation,
from the formula:
%%%%%%%%%%%%%%%%%%%%%%%%%%%%%%%%%%%%%%%%%%%%%%%%%%%%%%%%%%%%%%%%%%%%%%%%%%%%%
$$ {\cal C}_\Omega \simeq {\cal C}\equiv
\left(
{ {\cal C}_M \over {{\cal C}_R}^3 }
\right)^{1/2}.
\eqno (2.4) $$
%%%%%%%%%%%%%%%%%%%%%%%%%%%%%%%%%%%%%%%%%%%%%%%%%%%%%%%%%%%%%%%%%%%%%%%%%%%%%
As  can be seen from Table 1, for causal EOS Eq. (2.4)
reproduces the actual values of
${\cal C}_\Omega({\rm EOS})$ within better than 5\%. In the case
when one considers also non-causal EOS, the precision of
approximation (2.4) worsens to 7\%.

Relation (2.4) implies, that the angular velocity for mass shedding
is approximated by the formula
%%%%%%%%%%%%%%%%%%%%%%%%%%%%%%%%%%%%%%%%%%%%%%%%%%%%%%%%%%%%%%%%%%%%%%%%%%%%%%
$$ \Oma \simeq\left({G M_{\rm max}({\rm rot})\over
                            R_{\rm max}^3({\rm rot})}\right)^{1/2}~.
                                                               \eqno (2.5)$$
%%%%%%%%%%%%%%%%%%%%%%%%%%%%%%%%%%%%%%%%%%%%%%%%%%%%%%%%%%%%%%%%%%%%%%%%%%%%%%

The value of $\Oma$ can be thus well approximated by the
frequency of a particle in stable circular orbit at the equator
of a fictitious {\it non-rotating} star of mass equal to $M_{\rm max}({\rm
rot})$ and of the radius equal to the equatorial radius of the
maximally rotating configuration, $R_{\rm max}({\rm rot})$ (in
the case of non-rotating star the general relativistic formula
for the particle frequency is identical to the newtonian one,
see {\it e.g.} Misner, Thorne \& Wheeler 1973).

Up to now, we considered only {\it realistic} EOS of dense
matter.
It is instructive to study  also the case of polytropic configurations with
an equation of state in the form $P=K {n_{\rm b}}^{\Gamma}$,
where $n_{\rm b}$ is baryon density of matter.
Extensive calculations of rotating polytropic models were
presented in Cook et al. (1992, 1994a).
 For a fixed value of $\Gamma$, static and rotating models
exhibit useful scaling properties with respect to change of $K$
(Cook et al. 1992, 1994a).
 Consequently, the relativistic parameter $x_{\rm s}$, and
 ${\cal C}_M({\rm EOS})$,
 ${\cal C}_R({\rm EOS})$ turn out to
be  functions of $\Gamma$ only; they are given in Table 2.
 For a fixed value of $\Gamma$, relations  (1.1), (2.2), and
(2.3) are thus exact, with  $\Gamma$-dependent values of
the numerical coefficients.

The realistic  EOS are not polytropes, and their local
adiabatic index, $\Gamma=(n_{\rm b}/P){\rm d}P/{\rm d}n_{\rm
b}$, depends on the density, $\Gamma=\Gamma(n_{\rm b})$. Two
specific examples of the density dependence of $\Gamma$ are
shown in Fig. 6. In both cases, the value of $\Gamma$ varies
within the relevant interval of $n_{\rm b}$ by more than $30\%$
of its maximum value. Clearly, such EOS cannot be represented by
 single polytropes. This explains, e.g.,  the lack of  monotonic trend
in the dependence of ${\cal C}_M({\rm EOS})$ on $x_{\rm s}({\rm
EOS})$ for realistic EOS, displayed in Fig. 5. Such a
non-monotonic, irregular behavior, characteristic of realistic
EOS, is to be contrasted with a monotonic dependence ${\cal
C}_M(x_{\rm s})$ for polytropes, Table 2. In spite of this
irregular behavior of  ${\cal C}_M({\rm EOS})$ for realistic
EOS, the values of  ${\cal C}({\rm EOS})$ for realistic EOS,
calculated from (2.4), show a clear trend for a monotonic
increase with $x_{\rm s}$. This results from a clear trend in
 ${\cal C}_R({\rm EOS})$ to decrease with $x_{\rm s}$, which --
magnified  by the third power within the bracket of formula
(2.4) -- dominates the ${\cal C}$ -- $x_{\rm s}$ relation.

%CHANGE: the paragraph below has been added
The trend for a monotonic increase of ${\cal C}_\Omega$ with
$x_{\rm s}$,  Fig. 1, can be well reproduced by a linear function
${\cal C}_\Omega^{\rm lin}(x_{\rm s})=
0.468 + 0.378x_{\rm s}$.
This function, obtained by a least squares fit to the points
shown in Fig. 1, reproduces
${\cal C}_\Omega({\rm EOS})$
(and
$\Omega_{\rm max}({\rm EOS})$,
if inserted in Eq. (1.1)) with
precision better than $1.5\%$, with typical relative error being
less than $1\%$.  It should be stressed, however, that such a
linear approximation
${\cal C}_\Omega^{\rm lin}(x_{\rm s})$
 is valid only within a rather narrow interval of $x_{\rm s}$,
characteristic of realistic EOS, $0.45<x_{\rm s}<0.7$. In
contrast to the one-parameter empirical formula with
${\cal C}_\Omega=0.67$,
${\cal C}_\Omega^{\rm lin}(x_{\rm s})$ does not work
 for the EOS
of the free neutron gas (see HSB).

\section{THE SLOW ROTATION APPROXIMATION}

Weber and Glendenning (1991, 1992) tried to derive the empirical
formula,  equation (1.1), by using the slow rotation approximation of
the Einstein equations.
 One should notice however that a neutron star rotating with $\Oma$ may
not be really  considered to be a slow rotator (Hartle \& Thorne 1978).
Indeed, let us define a
dimensionless angular velocity $\Omega_*$,
%%%%%%%%%%%%%%%%%%%%%%%%%%%%%%%%%%%%%%%%%%%%%%%%%%%%%%%%%%%%%%%%%%%%%%%%%%%%%%
$$ \Omega_*^{~2} \equiv {\Omega}^2/\left({GM\over R^3}\right),
\eqno (3.1)$$
%%%%%%%%%%%%%%%%%%%%%%%%%%%%%%%%%%%%%%%%%%%%%%%%%%%%%%%%%%%%%%%%%%%%%%%%%%%%%%%
where $M$ is the mass of the static star, $R$ its radius and
${\Omega}$ is the angular velocity
measured by observers at infinity.  The assumption of the slow
rotation means that (Hartle 1967),
%%%%%%%%%%%%%%%%%%%%%%%%%%%%%%%%%%%%%%%%%%%%%%%%%%%%%%%%%%%%%%%%%%%%%%%%%%%%%%%
$$ \Omega_*^{~2} \ll 1. \eqno (3.2)$$
%%%%%%%%%%%%%%%%%%%%%%%%%%%%%%%%%%%%%%%%%%%%%%%%%%%%%%%%%%%%%%%%%%%%%%%%%%%%%%%
 From equation (1.1) with ${\cal C}_{\Omega}=0.67$ it follows that
%%%%%%%%%%%%%%%%%%%%%%%%%%%%%%%%%%%%%%%%%%%%%%%%%%%%%%%%%%%%%%%%%%%%%%%%%%%%%%%
$$ \Omega_{*\ \rm max}^{~2}\approx 0.45,   \eqno(3.3) $$
%%%%%%%%%%%%%%%%%%%%%%%%%%%%%%%%%%%%%%%%%%%%%%%%%%%%%%%%%%%%%%%%%%%%%%%%%%%%%%%
so that the use of slow-rotation approximation for maximally rotating
configurations requires some additional justification.
It is unlikely that this approximation will give accurate results
considering the fact that at the stellar surface the linear speed of
rotation is a significant fraction of the speed of light (Hartle \&
Thorne 1968):
%%%%%%%%%%%%%%%%%%%%%%%%%%%%%%%%%%%%%%%%%%%%%%%%%%%%%%%%%%%%%%%%%%%%%%%%%%%%
$$ {v_S \over c} = {1\over \sqrt 2} \left({x_s \over 1 - x_s}\right)^{1/2}
\eqno(3.4) $$
%%%%%%%%%%%%%%%%%%%%%%%%%%%%%%%%%%%%%%%%%%%%%%%%%%%%%%%%%%%%%%%%%%%%%%%%%%%%
For the maximally rotating models of SBGH  one gets typically
$v_S/c \approx 0.7$.
In the slow rotation approximation, one can obtain for
$\Omega_{\rm max}$ an expression of the form, which seems to be
similar to that of the empirical formula (1.1) (see also
Glendenning and Weber 1994):
%%%%%%%%%%%%%%%%%%%%%%%%%%%%%%%%%%%%%%%%%%%%%%%%%%%%%%%%%%%%%%%%%%%%%%%%%%%%%%%
$$ \Omega_{\rm max} =
{\cal C}_{\rm sr}\left ({GM_{\rm max}({\rm rot})
    \over R_{\rm max}^{~3}({\rm rot})}\right)^{1/2}
     + {\cal O}(\Omega_*^3), \eqno (3.5)$$
%%%%%%%%%%%%%%%%%%%%%%%%%%%%%%%%%%%%%%%%%%%%%%%%%%%%%%%%%%%%%%%%%%%%%%%%%%%%%%%
with
%%%%%%%%%%%%%%%%%%%%%%%%%%%%%%%%%%%%%%%%%%%%%%%%%%%%%%%%%%%%%%%%%%%%%%%%%%%%%%
$$ {\cal C}_{\rm sr} \approx \left [ 1
             + {{I}\over {MR^2}}  {{R_G}\over R}\left(1 - 2.5{I\over MR^2}
                \left({R\over R_G}\right)^4
                \left(1 -{R_G\over R}\right)Q^{'2}_2(u)\cdot
                \left(1-{{QMc^2}\over J^{2}}\right) \right)
		\right]^{-{1\over 2}}~.
                 \eqno (3.6)$$
%%%%%%%%%%%%%%%%%%%%%%%%%%%%%%%%%%%%%%%%%%%%%%%%%%%%%%%%%%%%%%%%%%%%%%%%%%%%%%%
Here $R_G$ is the gravitational radius,
%%%%%%%%%%%%%%%%%%%%%%%%%%%%%%%%%%%%%%%%%%%%%%%%%%%%%%%%%%%%%%%%%%%%%%%%%%%%%%%
$$ R_G = {{2GM}\over c^2}, \eqno (3.7)$$
%%%%%%%%%%%%%%%%%%%%%%%%%%%%%%%%%%%%%%%%%%%%%%%%%%%%%%%%%%%%%%%%%%%%%%%%%%%%%%%
$J$ is the angular momentum, $I$ the moment of inertia, $Q$ is
the quadrupole moment of the rotating configuration,   and
$Q^{'2}_{2}(u)$  is the derivative, with respect to
$u=1-R/R_G$, of the associated Legendre function of the second kind.
%%%%%%%%%%%%%%%%%%%%%%%%%%%%%%%%%%%%%%%%%%%%%%%%%%%%%%%%%%%%%%%%%%%%%%%%%%
%$$   Q^{'2}_{2} ({ u}) = 3{ u}
%           \log{{ u} + 1 \over { u} -1}
%                      -{3{ u}^4 - { u}^2 -6{ u} + 8 \over
%                      \left({ u}^2 - 1 \right)^2}~,
%                      \eqno(3.8)  $$
%%%%%%%%%%%%%%%%%%%%%%%%%%%%%%%%%%%%%%%%%%%%%%%%%%%%%%%%%%%%%%%%%%%%%%%%%%%
 The quantities $M$ and $R$ are those for the
maximally rotating configuration, but -- within our
approximation ---  we can as well replace them by
$M_{\rm max}({\rm stat})$, $R_{\rm max}({\rm stat})$.

Equations (3.5) and (3.6) give $\Omega_{\rm max}$ in terms of the mass,
 equatorial radius, moment of inertia, angular momentum
 and the quadrupole moment of the maximally {\it
rotating} configuration. At first glance (and neglecting terms
$\sim \Omega_*^{~3}$ and higher),
expression (3.5 -- 3.6) may seem to be  similar to
 empirical formula (1.1).
It has been shown by Abramowicz and Wagoner (1978) and recently confirmed
by Ravenhall and Pethick (1994) that moments of inertia,
expressed in the units of $MR^2$,  are  -
to a good approximation - functions of
only $x=R/R_G$.
 Moreover, results of SBGH show that  ${QMc^2}/ J^{2}$
is a decreasing
function of $x_s$. Expression (3.5) seems thus to possess
similar properties as
the empirical formula. Unfortunately {\it it is not} the
empirical  formula since it involves $M_{\rm max}$ and $R_{\rm
max}$ of {\it rotating} configurations, and therefore it
corresponds to some approximation of eq. (2.5) and not to eq. (1.1).
Further expansions of $M_{\rm max}({\rm rot})$ and $R_{\rm
max}({\rm rot})$ in $\Omega_*$ are not justified - in view of
the large value of this parameter (notice that rotation increases
the value of $R_{\rm max}$ by some 30\%).
 So, only some external  ``empirical input" (such as
assumption of a  ``typical" effect of rotation on $M_{\rm max}$,
$R_{\rm max}$, and on eccentricity of rotating star, made in
Weber \& Glendenning (1992)) can lead to an expression of type
(1.1). A consistent application of the slow rotation
approximation cannot reproduce the empirical formula for
$\Omega_{\rm max}$.
 This is
not very surprising if one considers that one should get to the 8-th
order in $\Omega_*$  to get a precision of 4\%, characteristic
of empirical formula.

\section{DISCUSSION AND CONCLUSIONS}

The proportionality constant appearing in the empirical formula for
$\Omega_{\rm max}$ for realistic EOS of dense matter,
 is in fact a function of the
relativistic parameter $x_s$. In the range of parameters
describing maximally rotating configurations with causal equations
of state ${\cal C}_{\Omega}$ is a rather slowly varying function
of $x_s$, which results in a high precision of empirical formula
with an appropriate choice of a universal proportionality
constant.
We found universal relations connecting maximal masses and corresponding radii
of static neutron stars,
calculated for realistic EOS of dense matter,
 with those of maximally rotating configurations.
The empirical  formula follows from those relations.

Although the slow rotation approximation allows one to reproduce some
of the properties of the empirical formula this approximation is
not appropriate for maximally rotating configurations.

%In any case one should first try to understand the origin of the
%universal mass and radius relations found in the present work.

It is a great pleasure to thank Silvano Bonazzola for his role in
stimulating this research. We are grateful to him, Eric Gourgoulhon,
Marcelo Salgado and R\'emi Hakim for very useful discussions.
We thank the second referee of this paper for useful suggestions and comments.
 This research was partially supported by the Polish State Committee
for Scientific Research  (KBN) grant, and by  PICS/CNRS no. 198
``Astronomie Pologne".

\vfill\eject
\begin{table}
\begin{tabular}{llllll}
\multicolumn{6}{c}{TABLE 1}\\
&&&&\\
\multicolumn{6}{c}{REALISTIC EQUATIONS OF STATE}\\
&&&&\\
\hline\hline
&&&&\\
Equation of state & $x_{\rm s}$ & ${\cal C}_{\Omega}$ & ${\cal C}_{M}$ &
${\cal C}_{R}$ & ${\cal C}$ \\
&&&&\\
\hline
&&&&\\
Glendenning 1985 ``case 2" \hfill
   &   0.467  &  0.648    &   1.1741     &  1.3960     &  0.657 \\
   &&&&\\
Glendenning 1985 ``case 1"
  &   0.480  &  0.659    &   1.1783     &  1.3794     &  0.670   \\
  &&&&\\
Glendenning 1985 ``case 3"
  &   0.516  &  0.658    &   1.1746     &  1.3740     &  0.673 \\
  &&&&\\
Diaz Alonso 1985 model II
   &   0.524  &  0.666    &   1.1704     &  1.3575     &  0.684   \\
   &&&&\\
Weber et al. 1991 $\Lambda^{00}_{\rm Bonn}+{\rm HV}$
  &   0.533  &  0.661    &   1.1983     &  1.3786     &  0.676   \\
  &&&&\\
Bethe \& Johnson 1974 model IH
   &   0.554  &  0.672    &   1.1597     &  1.3409     &  0.693   \\
   &&&&\\
Wiringa et al. 1988 model ${\rm UV14+TNI}$
   &   0.569  &  0.687    &   1.1826     &  1.3277     &  0.711   \\
   &&&&\\
Pandharipande 1970
    &   0.577  &  0.684    &   1.1633     &  1.3226     &  0.709   \\
    &&&&\\
Haensel et al. 1980 ``model 0.17"
  &   0.614  &  0.704    &   1.2138     &  1.3103     &  0.7345   \\
  &&&&\\
  Friedman \& Pandharipande 1981$^a$
   &   0.619  &  0.701    &   1.1807     &  1.3145     &  0.721  \\
   &&&&\\
Wiringa et al. 1988 model ${\rm UV14+UVII}~^a$
  &   0.656  &  0.719    &   1.1824     &  1.2690    &  0.761   \\
  &&&&\\
Wiringa et al. 1988 model ${\rm AV14+UVII}~^a$
 &   0.667  &  0.722    &   1.1904     &  1.2671     &  0.765   \\
 &&&&\\
 \hline
\end{tabular}
\parindent 0pt
\vskip3mm
{}~~~~~~~ $^a$ Non-causal within central cores of massive neutron stars
\end{table}

\begin{table}
\begin{center}
\begin{tabular}{cccccc}
\multicolumn{6}{c}{TABLE 2}\\
&&&&\\
\multicolumn{6}{c}{POLYTROPIC EQUATIONS OF STATE}\\
&&&&\\
\hline\hline
&&&&\\
 $\Gamma$ & $x_{\rm s}$
 & ${\cal C}_\Omega$ & ${\cal C}_M$ & ${\cal C}_R$& ${\cal C}$\\
 &&&&\\
\hline
&&&& \\
&&&& \\
2.00 & 0.438 & 0.624 & 1.1504 & 1.4242 & 0.631 \\
&&&& \\
2.25 & 0.509 & 0.640 & 1.1766 & 1.4032 & 0.652\\
&&&& \\
2.50 & 0.563 & 0.660 & 1.1961 & 1.3745 & 0.679\\
&&&& \\
2.75 & 0.605 & 0.6745 & 1.2111 & 1.3540 & 0.6985\\
&&&& \\
3.00 & 0.637 & 0.6923 & 1.2225 & 1.3288 & 0.723\\
&&&&\\
\hline
\end{tabular}
\end{center}
\end{table}

\centerline{\bf FIGURE CAPTIONS}
\parindent 0pt
\medskip
{\bf Fig. 1.} Proportionality constant
 ${\cal C}_\Omega ({\rm EOS})$ for twelve
realistic  EOS versus relativistic parameter $x_{\rm s}$. Filled
circles for causal EOS, open circles for EOS which are
non-causal within massive neutron star models.
\vskip 3mm

{\bf Fig. 2.} Mass of  maximally rotating neutron stars versus
maximum mass of non-rotating models for realistic EOS. Filled
circles for causal EOS, open circles for EOS which are
non-causal within massive neutron star models. Straight line
gives the  best fit to exact results obtained for causal EOS.
\vskip 3mm

{\bf Fig. 3.} Equatorial radius
 of  maximally rotating neutron stars versus
the radius of the  static maximum mass configurations
 for realistic EOS. Filled
circles for causal EOS, open circles for EOS which are
non-causal within massive neutron star models. Straight line
gives the  best fit to exact results obtained for causal EOS.
\vskip 3mm

{\bf Fig.4.} Proportionality constant ${\cal C}_R ({\rm EOS})$ for twelve
realistic  EOS versus relativistic parameter $x_{\rm s}$. Filled
circles for causal EOS, open circles for EOS which are
non-causal within massive neutron star models. Dashed horizontal
line corresponds to the  best fit to exact results for causal EOS.
\vskip 3mm

{\bf Fig.5.} Proportionality constant ${\cal C}_M ({\rm EOS})$ for twelve
realistic  EOS versus relativistic parameter $x_{\rm s}$.
 Notation as in Fig. 4.
 \vskip 3mm

{\bf Fig. 6.} Adiabatic index, $\Gamma$, versus baryon density,
$n_{\rm b}$, for  AV14 + UVII (solid line) and UV14 + TNI
(dashed line) EOS. The calculation of $\Gamma$ has been based on
the polynomial fit of Kutschera \& Kotlorz (1993) to the
tabulated EOS of Wiringa et al. (1988).
\vskip 3mm

\vfill
\end{document}